\documentclass[reprint,amsmath,amssymb,prl,aps,floatfix,superscriptaddress,noeprint]{revtex4-1}
\usepackage{graphicx}
\usepackage{bm}
\usepackage{upgreek}
\usepackage{hyperref}
\usepackage[mathlines]{lineno}

\begin{document}

\author{Sander A. Mann}
\affiliation{Department of Electrical and Computer Engineering, The University of Texas at Austin, Austin, TX, USA}

\author{Dimitrios L. Sounas}
\affiliation{Department of Electrical and Computer Engineering, The University of Texas at Austin, Austin, TX, USA}

\author{Andrea Al{\`u}}
 \email{aalu@gc.cuny.edu}
 \affiliation{Department of Electrical and Computer Engineering, The University of Texas at Austin, Austin, TX, USA}
\affiliation{Photonics Initiative, Advanced Science Research Center, City University of New York, New York 10031, USA}
\affiliation{Physics Program, Graduate Center, City University of New York, New York 10016, USA}
\affiliation{Department of Electrical Engineering, City College of The City University of New York, New York 10031, USA}

\date{\today}

\title{Nonreciprocal Cavities and the Time-Bandwidth Limit}
\begin{abstract}\noindent
The time-bandwidth limit inherently relates the lifetime of a resonance and its spectral bandwidth, with direct implications on the maximum storage time of a pulse versus its frequency content. It has been recently argued that nonreciprocal cavities may overcome this constraint, by breaking the strict equality of their incoupling and outcoupling coefficients. Here, we generally study the implications of nonreciprocity on resonant cavities and derive general relations, stemming from microscopic reversibility, that govern their dynamics. We show that nonreciprocal cavities do not provide specific advantages in terms of the time-bandwidth limit, but they may have other attractive properties for nanophotonic systems.
\end{abstract}

\maketitle

The demands of integrated photonic circuits have motivated a large body of research with the objective of scaling down and integrating crucial optical components for information processing \cite{Reed2010,Sounas2017,Ayata2017,Lira2012,Green2007,Bi2011,Phare2015,Xu2005,Shen2015,Manolatou1999,Grote2013,Feng2012,Sounas2014}. A common approach to enable small footprints is to use resonant cavities \cite{Bi2011,Phare2015,Xu2005,Shen2015,Manolatou1999,Grote2013,Feng2012,Sounas2014}, which can dramatically enhance light-matter interactions by storing light over the resonance lifetime. This reduced footprint, however, comes at the cost of operational bandwidth: a single resonant cavity adheres to a strict correspondence between its resonance bandwidth $\Delta \omega$ and its lifetime $\Delta t$ \cite{SI}: $\Delta t \Delta \omega = 2$. As a relevant example in nanophotonics, resonators are frequently used to impart delays on optical pulses, given their ability to temporarily store light. The strict relation between lifetime and bandwidth of a cavity implies a trade-off between the spectral bandwidth of the pulse that can be stored and the temporal delay that can be imparted on it. Larger structures (e.g., arrangements of multiple cavities \cite{Yariv1999}) may partially relax this bound, but still follow an analogous trade-off between the delay-bandwidth product and the overall footprint  \cite{Miller2007}. Recent approaches to overcome this limitation have relied on time-varying systems \cite{Yanik2004,Xu2007,Minkov2017} or on nonlinearities \cite{Lannebere2015}, but efficiently implementing these schemes often becomes technologically challenging, or imposes limitations in their operation. 

A thought-provoking proposal to alleviate the strict correspondence between bandwidth and lifetime of linear, time-invariant structures has been recently been put forward, based on breaking reciprocity with a static magnetic bias \cite{Tsakmakidis2017}. The authors argue that the supported bandwidth $\Delta \omega$ scales with the rate at which energy enters the system, while the lifetime is inversely proportional to the rate at which energy exits it. These rates are known to be equal in reciprocal cavities \cite{Haus1984}, resulting in the mentioned time-bandwidth relation, but Ref.~\cite{Tsakmakidis2017} suggests that nonreciprocal cavities may support different input and output rates, and thus can surpass this limit. While this idea would have a large impact on many photonic applications relying on resonances, it raises concerns regarding its thermodynamic validity \cite{Tsang2018}: unequal input and output rates violate energy conservation in a passive system.

Inspired by this proposal, in the following we develop a general temporal coupled-mode theory broadly describing the dynamics of nonreciprocal cavities, and show that incoming and outgoing rates are always strictly related through a number of identities that generalize the known relations in reciprocal systems. For example, we prove that in nonreciprocal cavities the total rates must always be equal in magnitude. We also show that time-bandwidth product is strictly limited in any linear, time-invariant cavity, independent of whether it is reciprocal or not, and provide numerical demonstrations. We also derive bounds for the coupling coefficients of nonreciprocal cavities, and design cavities operating at these bounds.  

We start our discussion by studying a general resonant cavity with complex mode amplitude a connected to n ports, with incoming complex amplitudes $\mathbf{s}_+=(s_{+,1},s_{+,2},\cdots,s_{+,n} )^\intercal$. Such a system is described by the equation of motion \cite{Haus1984,Suh2004}

\begin{equation}\label{tcmt}
\frac{da}{dt} = (i\omega - \gamma_r -\gamma_i)a+\mathbf{k}^\intercal \mathbf{s}_+
\end{equation}
where $\mathbf{k}=(k_1,k_2,\cdots,k_n )^\intercal$ is the vector of input coupling coefficients, $\omega_0$ is the eigenfrequency, $\gamma_r$ is the radiation loss rate into the ports, and $\gamma_i$ is the intrinsic loss rate due to absorption inside the cavity. The amplitudes are normalized such that $\lvert a \rvert^2$ is the stored energy in the cavity and $\lvert s_{+,i} \rvert^2$ is the incoming power at port $i$. The amplitude of the reflected modes is given by

\begin{equation}\label{reflection}
\mathbf{s}_- = \mathbf{C} \mathbf{s}_+ + \mathbf{d}a
\end{equation}
Here, $\mathbf{C}$ is the direct reflection coefficient, which is unitary in the absence of loss, and $d$ is the output coupling coefficient from the cavity mode into the ports. Power conservation ensures that $2\gamma_r = \mathbf{d}^\dagger \mathbf{d}$ \cite{Suh2004}. In addition, by assuming time-harmonic excitation ($s_+ = s_{+,0} \exp{(i\omega t)}$) we find from Eq. \ref{tcmt}:

\begin{equation}\label{amplitude}
a(\omega) = \frac{\mathbf{k}^\intercal \mathbf{s}_+}{i(\omega-\omega_0) + \gamma_r +\gamma_i}
\end{equation}
Eqs. \ref{tcmt}-\ref{amplitude} and the relation $2\gamma_r = \mathbf{d}^\dagger \mathbf{d}$ apply to both reciprocal and nonreciprocal cavities. In addition, reciprocity dictates that $\mathbf{d}=\mathbf{k}$ and $\mathbf{C}\mathbf{d}^*=-\mathbf{d}$ (asterisk denotes complex conjugate) \cite{Suh2004}, which follow from time-reversal symmetry. However, if the cavity is biased with a quantity odd symmetric under time-reversal, these two identities do not apply, as the time-reversed system is no longer equivalent to the original one. In the following, we therefore denote the coefficients in the time-reversed scenario with a tilde, as $\mathbf{\tilde{C}}$, $\mathbf{\tilde{d}}$ and $\mathbf{\tilde{k}}$. As a first important result of this work, we show that, if the resonant frequency is not perturbed by a time-reversal operation, the following relationships necessarily hold for both reciprocal and nonreciprocal linear time-invariant systems (proofs are provided in \cite{SI}):

\begin{subequations} \label{identities}
\begin{equation}
\mathbf{\tilde{C}} = \mathbf{C}^\intercal
\end{equation}
\vspace{-0.8cm}
\begin{equation}
\mathbf{\tilde{d}} = \mathbf{k}
\end{equation}
\vspace{-0.8cm}
\begin{equation}
\mathbf{\tilde{k}} = \mathbf{d}
\end{equation}
\vspace{-0.8cm}
\begin{equation}
\mathbf{\tilde{\gamma}}_r = \mathbf{\gamma}
\end{equation}
\vspace{-0.8cm}
\begin{equation}
\mathbf{C}^\intercal\mathbf{d}^* = -\mathbf{k}
\end{equation}
\vspace{-0.8cm}
\begin{equation}
\mathbf{k}^\dagger \mathbf{k} = \mathbf{d}^\dagger \mathbf{d}
\end{equation}
\end{subequations}
Here the superscripts $\intercal$ and $\dagger$ denote the regular and Hermitian transpose. These relationships are markedly different from the regular expressions for reciprocal systems \cite{Suh2004,note1}, particularly because they involve the original system \emph{and} its time-reversed counterpart (which, as we will show, can differ strongly in its response). This is a consequence of microscopic reversibility \cite{Casimir1945}: while the system itself is not time-reversal symmetric, global time-reversal invariance still applies, and the original and time-reversed systems are necessarily related.

Eq.~\ref{identities}f is the fluctuation-dissipation relation, relating dissipation ($\mathbf{d}$) to how the cavity responds to external fluctuations ($\mathbf{k}$) \cite{SI,Tsang2018,Gardiner2003}. Together with Eq.~\ref{identities}e, this relation is arguably the most important result for the following discussions, as they put stringent restrictions on the achievable asymmetry in the coupling coefficients: while it is possible to achieve $d_i \neq k_i$ at any individual port, the positive-definite norm of $\mathbf{k}$ and $\mathbf{d}$ has to be equal (see Eq.~\ref{identities}f). Finally, we note that Eqs.~\ref{identities} readily reduce to the known identities for reciprocal cavities \cite{Suh2004}, if we assume that the system is time-reversal symmetric, i.e. $\mathbf{\tilde{d}} = \mathbf{d}$, $\mathbf{\tilde{k}} = \mathbf{k}$, and $\mathbf{C} = \mathbf{C}^\intercal$.

\begin{figure}[t]
\begin{center}
\includegraphics{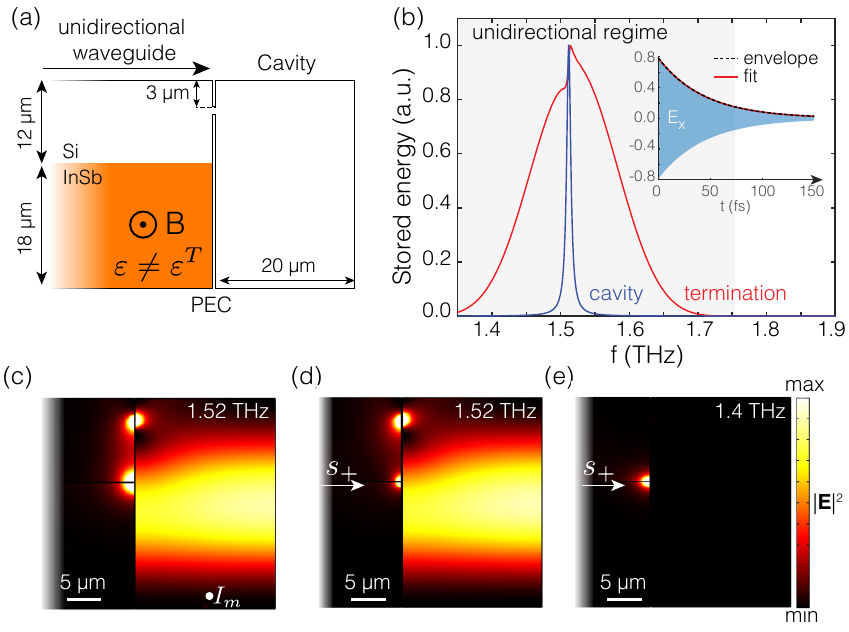}
\caption{(a) Schematic of the waveguide/cavity geometry under consideration (see [24] for additional details). (b) Fourier transform of $E_x$ inside the cavity at the center (blue) and outside of the cavity at the termination (red), with the unidirectional bandwidth of the waveguide shown in grey. Inset: ring down of the electric field inside the cavity, in perfect agreement with the line width obtained from the spectrum. (c) Electric field intensity near the cavity at 1.52 THz when excited from inside with a magnetic line source $I_m$, highlighting the dissipative mechanism at the wedge. (d,e) Electric field intensity near the cavity at 1.52 THz (d) and at 1.4 THz (e) when excited from the port. In these cases the wedge mode is also clearly visible. \vspace{-0.5cm}} 
\label{fig1}
\end{center}
\end{figure}

Eqs. (\ref{identities}) have important general consequences in the context of nonreciprocal cavities \cite{Yu2008,Liu2015}. As a relevant example, consider a nonreciprocal waveguide supporting a unidirectional mode, \emph{i.e.}, no backward channel into which energy can be reflected, and terminated into a cavity, similar to geometries studied in \cite{Luukkonen2012,Hadad2013} and shown schematically in Fig.~\ref{fig1}a. For the nonreciprocal waveguide, we use the interface between a magnetized magneto-optic semiconductor and a dielectric material \cite{Yu2008,Brion1972,Camley1987,Davoyan2013}, which for the combination InSb and Si is known to support a unidirectional surface wave around 1.5 THz \cite{Tsakmakidis2017,Shen2015b,Shen2015a} (see all geometry details in \cite{SI}). The waveguide is terminated onto a resonant lossless rectangular cavity (20 by 30 µm in size), enclosed by perfect electric conducting (PEC) walls, connected through a small aperture (0.5 $\upmu$m).

We use a home-built FDTD code \cite{SI} to excite the unidirectional waveguide with a broadband pulse ($f_0$=1.5 THz, $\Delta f$=0.16 THz), and record the fields inside the cavity as a function of time. Fig.~\ref{fig1}b shows the Fourier transform of the electric field induced at the center of the cavity (blue line), showing a sharp Lorentzian resonance at 1.52 THz, with linewidth $2\gamma =  45.2 \times 10^9$ rad/s. When exciting the cavity from the interior, we find exactly the same decay rate $\gamma=22.6 \times 10^9$ rad/s by examining the field decay (inset of Fig.~\ref{fig1}b). This should not be surprising, as Eq.~(\ref{amplitude}) applies independent of whether the system is reciprocal or not, and ensures that the bandwidth of any linear cavity is solely determined by the decay rate $\gamma=\gamma_r+\gamma_i$, \emph{i.e.}, by the internal loss and the outcoupling coefficients $\mathbf{d}^\dagger \mathbf{d}/2=\gamma_r$. This is supported by the observation that the fields at the termination outside the cavity have a much larger bandwidth (Fig.~\ref{fig1}b, red line), equal to the bandwidth of the incident pulse, but that these fields cannot be forced into the cavity. In other words, as a first important consequence of the previous theory, the time-bandwidth product of any linear cavity satisfies $\Delta t \Delta \omega = 2$, and it is controlled only by the output coupling coefficient $d$. The incoupling coefficient $k$, according to Eq.~(\ref{amplitude}), only controls the amount of stored energy in the resonator.

More surprising is that the cavity has a finite decay rate, despite the fact that it is both lossless ($\gamma_i=0$) and connected to a waveguide that does not support backward modes ($\gamma_r=0$). While the decay process is not immediately apparent, it is consistent with Eq.~\ref{identities}f, which requires that if the cavity can be excited it must also be able to decay to maintain equilibrium. This paradox can be resolved by inspecting the electric field distribution during decay (without an external excitation), shown in Fig.~\ref{fig1}c. We notice a strong hot spot at the wedge formed by the Si/InSb waveguide and the PEC termination which, given the finite material loss in InSb, dissipates all incoming energy and thus sustains the cavity decay. We also observe this hotspot when exciting from the waveguide port, both on- and off-resonance (Figs.~1d,e) \cite{Tsakmakidis2017,Shen2015a,Ishimaru1962,Chettiar2014,Marvasti2017}. Since both the cavity and the input port can excite this wedge mode, they interfere in the process of dissipation (as evidenced in the bump through the resonance in the red curve in Fig. 1b). As a result, the wedge mode cannot be treated as a simple internal loss process, as $\gamma_i$, but needs to be treated as an additional output port \cite{SI}. Then, writing $\mathbf{k}^\intercal=(k_r,k_w)$ and $\mathbf{d}^\intercal=(d_r,d_w)$ for the input and output coefficients for the radiation from/into the waveguide and wedge, respectively, we find that $d_r=0 \neq k_r $ is permitted, while $\mathbf{k}^\dagger \mathbf{k} = \mathbf{d}^\dagger \mathbf{d}$ is simultaneously satisfied. In other words, this means that while the input and output coefficients may differ at each individual port, but the total input and output rates must always be equal. It follows that in the special case of a one-port system, like the one generally described in \cite{Tsakmakidis2017}, the input and output coefficients must necessarily be equal in magnitude, but they may still differ in phase.

One may wonder what happens in the limit in which material loss in the unidirectional waveguide is zero, for which the wedge mode is expected to become non-dissipative. This problem has been extensively discussed in the literature (see \cite{Ishimaru1962,Lax1955,Kales1956,Seidel1957,Bresler1960} for a selection), and has been referred to as the ``thermodynamic paradox'', since it was believed to produce an inconsistency between Maxwell's equations and thermodynamics. The paradox was resolved by Ishimaru [36], who pointed out that the field distribution at the termination of a unidirectional waveguide is non-integrable if assumed lossless, and it sustains finite absorption even in the limit of vanishing material loss. A lossless terminated one-way waveguide is actually an ill-posed boundary-value problem, as Maxwell's equations do not guarantee a unique solution in the ideal lossless scenario \cite{Harrington2001}. The correct solution is found in the limit of vanishing loss, for which the dissipation rate remains finite. Aside from the wedge mode in this terminated unidirectional waveguide \cite{Tsakmakidis2017,Shen2015a,Ishimaru1962,Chettiar2014,Marvasti2017}, similar singularities that support finite absorption in the limit of vanishing material loss can be found in other extreme electromagnetic systems \cite{Silveirinha2007,Estakhri2013}, and are a reminder that ideal lossless scenarios should be considered an artifact in electromagnetics and may lead to singularities or non-unique solutions.

While Eqs.~\ref{tcmt}-\ref{identities} strictly prove that the time-bandwidth product of a linear time-invariant cavity is always equal to 2, and that therefore it is impossible to force broadband fields into a long-lived resonance, independent of reciprocity, Refs. \cite{Tsakmakidis2017,Shen2015a,Ishimaru1962,Chettiar2014,Marvasti2017} do demonstrate broadband focusing of photons in an ultra-small volume near the termination, whose decay rate is unrelated to, and can be much slower than, the excitation. Consistently, our simulations confirm focusing of incident photons at the wedge (Figs.~1d,e) over a wide range of frequencies (Fig.~1b, red line), irrespective of the properties of the cavity resonance. We stress however that this broadband focusing is not directly a consequence of nonreciprocity: adiabatically tapered terminated plasmonic waveguides \cite{Stockman2004,Pile2006,Verhagen2008}, which slowly focus the incoming fields towards an apex, perform the same function. The benefit of applying nonreciprocity is that the termination is automatically matched due to the absence of a backwards mode in the waveguide \cite{Luukkonen2012,Hadad2013}, relaxing the need for a carefully designed adiabatic transition that minimizes reflections.

After having demonstrated that nonreciprocity is not beneficial to break the trade-off between lifetime and bandwidth in linear, time-invariant cavities, in the following we will explore to what degree the input and output coupling may be made different in nonreciprocal cavities, and what functionalities can be enabled by such asymmetry. Achieving asymmetry in the coupling coefficients is important for \emph{e.g.} circulators \cite{Pozar2012} and unidirectional heat transfer \cite{Zhu2016,Zhu2018}. In the previous example, the absence of a backwards mode ensured $d_r=0$. However, we will now show that, even if there is a backwards mode, the input and output coefficients at a given port differ significantly in nonreciprocal cavities. To do so, we examine the same system as in Fig.~\ref{fig1}, but now excited in the bidirectional regime at frequencies just below 1.25 THz. We therefore tune the cavity to 1.24 THz, by increasing the cavity size to 35.4 by 20 $\upmu$m (see  \cite{SI} for geometry details). With full-wave simulations we retrieve the complex cavity and reflection amplitudes, and through a fitting procedure we obtain $k_r=(2.65+0.308i) \times 10^{-4} \sqrt{\text{rad/s}}$ and $d_r=(0.667+2.14i)\times 10^{-4} \sqrt{\text{rad/s}}$ (see \cite{SI} for spectra and fits). As expected, now $d_r \neq 0$  because of the presence of a propagating backward mode, but the two magnitudes are shown to be significantly different due to the presence of the wedge mode. When operating just below the unidirectional gap the system is still strongly asymmetric (in this case the forward and backward effective indices are 4.47 and 9.92 respectively), and it is to be expected that, as the asymmetry reduces, $k_r$ and $d_r$ will also approach each other.

\begin{figure}[t]
\begin{center}
\includegraphics{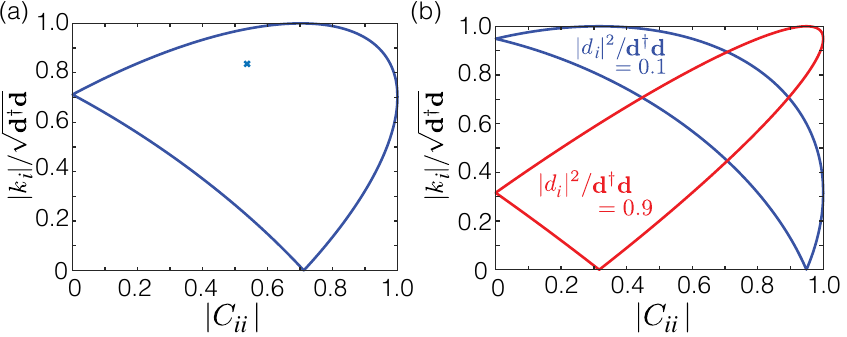}
\caption{(a) Bounds on $\lvert k_i \rvert$ as function of $\lvert C_{ii} \rvert$ for one of the cavities discussed in the main text, which has  $\lvert d_i \rvert^2/ \mathbf{d}^\dagger \mathbf{d}=0.53$. (b) Bounds in the cases that $\lvert d_i \rvert^2/ \mathbf{d}^\dagger \mathbf{d}=0.1$ and $\lvert d_i \rvert^2/ \mathbf{d}^\dagger \mathbf{d}=0.9$. \vspace{-0.5cm}} 
\label{fig2}
\end{center}
\end{figure}

While the input and output coefficients at the input port may be made different, both in magnitude and phase, Eq.~\ref{identities}f requires the total rates of the cavity must be the same, and in the system under consideration this is guaranteed by the wedge mode, which balances out any asymmetry in $k_r$ and $d_r$. Interestingly, this is, however, not the only requirement: the direct pathway places an additional bound on these coefficients through Eq.~\ref{identities}e. For a general two-port system with $\mathbf{k}^\intercal=(k_i,k_j)$ and $\mathbf{d}^\intercal=(d_i,d_j)$,  we find using the Cauchy-Schwarz inequality:

\begin{multline}\label{ineq}
\lvert ( \lvert C_{ii} d_i \rvert - \sqrt{1-\lvert C_{ii} \rvert^2})\rvert \leq \lvert k_i \rvert  \leq \\
 \lvert C_{ii} d_i \rvert + \sqrt{1-\lvert C_{ii} \rvert^2}
\end{multline}
where $\sqrt{1-\lvert C_{ii} \rvert^2} =\lvert C_{ij} \rvert^2$ due to power conservation. An example of this bound is shown in Fig.~\ref{fig2}a as a function of $\lvert C_{ii} \rvert$ for $\lvert d_i \rvert^2/ \mathbf{d}^\dagger \mathbf{d}=0.53$ ($\lvert d_i \rvert^2/ \mathbf{d}^\dagger \mathbf{d}$ quantifies the fraction of radiated power flowing towards port $i$ as the cavity decays). This specific example corresponds to the previously discussed bidirectional system, where the subscript $i$ indicates radiation into/from the port (e.g. $d_i=d_r$). The cross in the figure highlights the magnitude of $k_i$ and $C_{ii}$ for this geometry at 1.24 THz, falling within the bounds delimited by the solid line. The shape of the bound on $k_i$ strongly depends on $d_i$ and $d_j$: Fig.~2b displays similar bounds for $\lvert d_i \rvert^2/ \mathbf{d}^\dagger \mathbf{d}=0.1$ and $\lvert d_i \rvert^2/ \mathbf{d}^\dagger \mathbf{d}=0.9$. In the system of Fig.~\ref{fig1} he absence of a backward mode dictates that both $d_i=0$ and $C_{ii}=0$, in which case Eq.~\ref{ineq} becomes $\lvert k_i \rvert = \lvert d_j \rvert$: as expected, power can only enter the cavity at the rate in which it is dissipated by the wedge mode or other output port. Similarly, if $\lvert C_{ii} \rvert =1$, which implies that the two ports are not directly connected, Eq.~\ref{ineq} requires that $\lvert k_i \rvert = \lvert d_i \rvert$, and thus input and output coefficients for each port can be nonreciprocal only in phase, as in a gyrator. This is a consequence of the fact that the ports only exchange power with the cavity itself, and hence cannot compensate for any asymmetry in its response.

\begin{figure}[t]
\begin{center}
\includegraphics{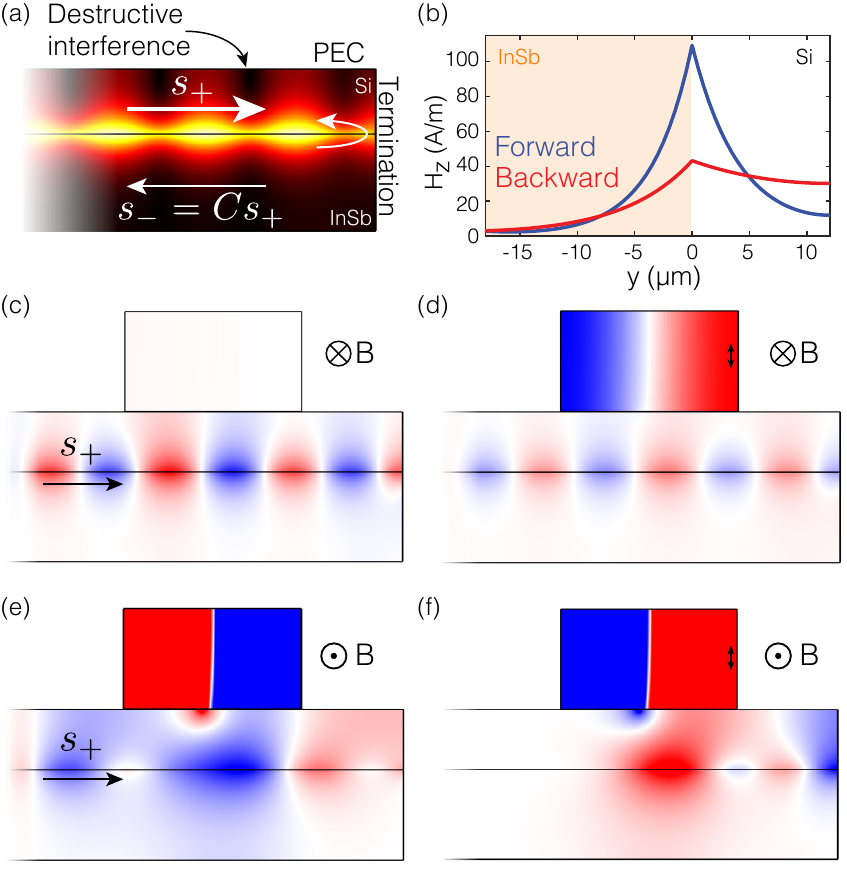}
\caption{(a) For the proper reflection amplitude, there are positions in a nonreciprocal waveguide with perfect destructive interference in the magnetic field, as visible in the magnetic field norm shown here (black is low and white is high field amplitude). (b) The required reflection amplitude $C=0.4$ is determined by the field ratio of the forward and backward power-normalized mode profiles. (c) A color plot of $\Re{(H_z)}$, demonstrating that a cavity with its opening at the position of destructive interference cannot be excited. (d) However, due to the nonreciprocal nature of the system, if the resonance is excited from inside the cavity it does decay into the port. (e,f) By reversing the direction of the biasing magnetic field, the cavity can be excited from the port but not decay into it. The color scale is clipped at 1/25$^{th}$ of the scale in (c,d) to enhance visibility of the fields in the waveguide. \vspace{-0.5cm}} 
\label{fig3}
\end{center}
\end{figure}

Having discussed the general bounds on the input and output coupling coefficients of nonreciprocal cavities, we now investigate an extreme condition allowed by the bounds in Eq.~\ref{ineq}, which may provide interesting functionalities for nanophotonic systems. We consider the scenario of a cavity connected to a bidirectional waveguide ($\lvert C_{ii} \rvert>0$) that cannot be excited from its input port ($k_i=0$), but that decays into it ($\lvert d_i \rvert>0$). To design such a cavity, it is sufficient to place its opening at a position with complete destructive interference between the forward and backward modes, as shown in Fig.~\ref{fig3}a. Since nonreciprocal waveguides have different forward and backward field profiles [50], we need to ensure that the backward mode has higher fields at the top PEC plate, and that the reflection coefficient $C_{ii}$ compensates for differences in magnetic field amplitude. As shown in Fig.~\ref{fig3}b, we achieve this condition by reverting the magnetic field bias with respect to the previous examples and optimizing of the reflectivity of the waveguide termination, using a dissipative surface impedance, to achieve $\lvert C_{ii} \rvert = 0.4$. The necessity of a reduced reflection coefficient is consistent with our previous finding, in Eq.~\ref{ineq}, that a nonreciprocal response in magnitude can only be achieved if $\lvert C_{ii} \rvert<1$.

We now find a locally vanishing magnetic field at the top PEC wall (Fig.~\ref{fig3}a), implying that a cavity coupled to the waveguide at that location cannot be excited. In Fig.~\ref{fig3}c, we place a cavity above the waveguide with its opening at the location of destructive interference, and a forward wave impinging from the channel indeed does not excite the cavity. If the waveguide were reciprocal, a zero in magnetic field at the cavity opening when excited from the port would imply that the cavity also cannot decay into the port. However, in the nonreciprocal case this is not so: if we excite the cavity from inside, we see it decay freely towards the output port (Fig.~\ref{fig3}d). This is a result of the fact that, due to nonreciprocity and the different profiles of the forward and backward modes of the waveguide, for excitation from inside the cavity these modes are excited with different amplitudes at the aperture.

Interestingly, as indicated by Eqs.~\ref{identities}c,d we can swap the values of $\mathbf{k}$ and $\mathbf{d}$ by reversing the direction of the magnetic field bias (which is equivalent to a time-reversal operation). This is shown in Figs.~\ref{fig3}e,f, which present the same structure but with a opposite magnetic bias. It can be seen that the cavity now can be efficiently excited from the port (Fig.~3e), but when the cavity is excited from inside it decays only towards the termination and get dissipated there, despite the presence of a backward propagating mode. It is interesting to point out that, in contrast to the example in Fig.~1, here $d_i=0$ while $C_{ii}>0$ --- even though there is an available backwards mode, the cavity cannot couple to it.

To conclude, in this paper we have presented a general theoretical framework describing the dynamics of nonreciprocal cavities. Our results show that, for single port systems, $\mathbf{k}$ and $\mathbf{d}$ can only differ in phase, and for $k_i$ and $d_i$ at any individual port to be different in magnitude at least one additional channel is necessary. This is consistent with the general principle that to realize a linear isolator it is necessary to have a loss channel \cite{Carlin1955,Gamo1959} or an additional radiative port, as in a circulator. For systems with multiple ports we have shown that the sums of all input and output rates are necessarily equal: $\mathbf{k}^\dagger \mathbf{k} = \mathbf{d}^\dagger \mathbf{d}$. This implies that non-reciprocal cavities still follow strict bounds on their input and output coefficients, but nonetheless can be employed to realize highly non-trivial phenomena, such as cavities that can be pumped only one-way and release the energy into a totally different channel. We have shown that under a reversal of the magnetic field bias the functionalities of such a system strictly reverse. Finally, we have shown that the bandwidth of a linear, time-invariant cavity is always inversely proportional to its decay rate, both in reciprocal and nonreciprocal systems. The decay rate of a cavity is solely determined by the internal loss and the outcoupling coefficients $\mathbf{d}$, not by the incoupling rate. Thus, it is not possible to force broadband fields into a long-lived resonance, independent of reciprocity, for which we have provided a numerical demonstration. Our results clarify claims that non-reciprocity may alleviate the strict limitations imposed by the trade-off between delay and bandwidth in photonic systems, and may help envisioning new efficient nonreciprocal components for, \emph{e.g.}, information processing, unidirectional transport, and thermal rectifiers. This work was supported by the Air Force Office of Scientific Research.

\bibliography{library.bib}

\begin{thebibliography}{54}
\bibitem [{not()}]{note1}%
  \BibitemOpen
  \href@noop {} {\bibinfo  {journal} {It is important to note a conceptual
  difference in coupled-mode theory when dealing with non-reciprocal systems:
  if $k_i$ corresponds to the coupling coefficient of a given forward mode to
  the resonance, then usually $d_i$ is the coupling coefficient to the backward
  version of the same mode. However, in nonreciprocal systems a subtler
  definition of these coefficients is required, as the waveguide might be
  unidirectional, or have largely different propagation properties in the two
  directions. Here, we assume (without loss of generality) that port $i$
  supports at least one forward or backward mode, and at most both. If there is
  no forward or backward mode, we set $k_i=0$ or $d_i=0$ respectively}\
  }\BibitemShut {NoStop}%
\end{thebibliography}

\end{document}


\author{Sander A. Mann}
\affiliation{Department of Electrical and Computer Engineering, The University of Texas at Austin, Austin, TX, USA}

\author{Dimitrios L. Sounas}
\affiliation{Department of Electrical and Computer Engineering, The University of Texas at Austin, Austin, TX, USA}

\author{Andrea Al{\`u}}
 \email{aalu@gc.cuny.edu}
 \affiliation{Department of Electrical and Computer Engineering, The University of Texas at Austin, Austin, TX, USA}
\affiliation{Photonics Initiative, Advanced Science Research Center, City University of New York, New York 10031, USA}
\affiliation{Physics Program, Graduate Center, City University of New York, New York 10016, USA}
\affiliation{Department of Electrical Engineering, City College of The City University of New York, New York 10031, USA}

\date{\today}

\title{Supporting Information: Nonreciprocal Cavities and the Time-Bandwidth Limit}

\maketitle

\tableofcontents

\section{The Time-Bandwidth Product}\noindent
Starting from Eq.~3 in the main text, the stored energy of the resonance is given by:

\begin{equation}\tag{S1}
\lvert a(\omega) \rvert^2 = \frac{ \lvert \mathbf{k}^\intercal \mathbf{s}_+ \rvert^2}{(\omega - \omega_0)^2 + \gamma^2}.
\end{equation}
The full-width, half-maximum of this Lorentzian curve is given by $2\gamma$. Similarly, from Eq.~1 in the main text, if follows that, if the cavity has a nonzero amplitude $a_0$ at $t=0$ without the presence of an incoming wave, it evolves in time as

\begin{equation}\tag{S2}
a(t) = a_0 \exp(i\omega_0)\exp(-\gamma t).
\end{equation}
Here, the lifetime of the resonance $\Delta t$ is given by the time it takes the resonance amplitude to reach $e^{-1}$ of its original value: $\Delta t = \gamma^{-1}$. Hence, taking the product of the lifetime and bandwidth, we find $\Delta \omega \Delta t = 2$.

\section{Proofs for nonreciprocal CMT identities}\noindent
Here we provide proofs for the identities shown in the main text, which we repeat here for clarity:

\begin{subequations} \label{identities}
\begin{equation}\tag{4a}
\mathbf{\tilde{C}} = \mathbf{C}^\intercal
\end{equation}
\vspace{-0.8cm}
\begin{equation}\tag{4b}
\mathbf{\tilde{d}} = \mathbf{k}
\end{equation}
\vspace{-0.8cm}
\begin{equation}\tag{4c}
\mathbf{\tilde{k}} = \mathbf{d}
\end{equation}
\vspace{-0.8cm}
\begin{equation}\tag{4d}
\mathbf{\tilde{\gamma}}_r = \mathbf{\gamma}
\end{equation}
\vspace{-0.8cm}
\begin{equation}\tag{4e}
\mathbf{C}^\intercal\mathbf{d}^* = -\mathbf{k}
\end{equation}
\vspace{-0.8cm}
\begin{equation}\tag{4f}
\mathbf{k}^\dagger \mathbf{k} = \mathbf{d}^\dagger \mathbf{d}
\end{equation}
\end{subequations}
Before proving these relationships, it is important to point out the consequence of a time-reversal operation on general amplitudes of the mode and cavity amplitudes: performing a time-reversal operation leads to $T: \mathbf{s}_+ \rightarrow \mathbf{s}_-^*$, $T: \mathbf{s}_- \rightarrow \mathbf{s}_+^*$, and $T: a \rightarrow a^*$, since it effectively conjugates the temporal exponent in each amplitude and reverses the direction of any vector (such as the propagation vector). In the following we provide the proofs not in order of Eq.~4 in the main text, but in an order that makes more sense with respect to interdependencies. These proofs are based on the assumption that the mode profile and frequency are not affected by the time-reversal operation, and for Eqs.~4b,c,e closely follow the derivations for the reciprocal identities \cite{Suh2004}.

\subsection{Eq.~4a: $\mathbf{\tilde{C}} = \mathbf{C}^\intercal$}\noindent
The first relationship can be proven simply by considering reflection strongly detuned from resonance, so that $\mb{s}_- = \mb{Cs}_+$. Under a time-reversal operation, we then find $\mb{s}_+^* = \mb{\tilde{C}}\mb{s}_-^*$. Taking the conjugate, left-multiplying by $(\mb{\tilde{C}^*})^{-1}$, and using that C is unitary if the direct pathway is lossless, we find $\mathbf{\tilde{C}} = \mathbf{C}^\intercal$.

\subsection{Eq.~4d: $\tilde{\gamma}_r = \gamma_r$}\noindent
To prove Eq.~4d, we again invoke the time-reversed scenario of a decaying cavity. Without input, $\mb{s}_+ = 0$, there are no reflections in the time-reversed case: $\mb{\tilde{s}}_- = \mb{\tilde{s}}_+^* = 0$. Also, as mentioned earlier, the incident signal in the time-reversed case is $\mb{\tilde{s}}_+ = \mb{\tilde{s}}_-^*$. Hence, we can write for Eq.~2 in the main text: $\mb{\tilde{C}}\mb{s}_-^*+\mb{\tilde{d}} a^* = 0$. Given that $\mb{s}_-^* = \mb{d}^* a^*$, we find $\mb{\tilde{C}}\mb{d}^*=-\mb{\tilde{d}}$, which, when invoking unitarity of $\mb{C}$, yields $\mb{d}^\dagger \mb{d} = \mb{\tilde{d}}^\dagger \mb{\tilde{d}}$, and thus Eq.~4d. 

\subsection{Eqs.~4b,c: $\mb{\tilde{d}} = \mb{k}$ and $\mb{\tilde{k}} = \mb{d}$}\noindent
To prove Eqs. 4b,c, consider a lossless cavity with an initial amplitude decaying into the ports, while $\mb{s}_+ = 0$. In this scenario, both $a$ and $\mb{s}_-$ decay exponentially with complex frequency $\omega_0 - i \gamma$. If we then reverse time, we excite the cavity with an exponentially growing wave with amplitude $\mb{\tilde{s}}_+ = \mb{s}_-^*$ and frequency $\omega_0+i\gamma$. Now, starting from the equation of motion $(i\omega - i\omega_0+\gamma)a = \mb{k}^\intercal\mb{s}_+$, we find for the time-reversed scenario at frequency $\omega = \omega_0 +i \tilde{\gamma}$:

\begin{equation}\tag{S3}
(-i(\omega_0+i \tilde{\gamma})+i\omega_0 + \gamma)a^* = \mb{\tilde{k}}^\intercal \mb{s}_-^*.
\end{equation}
Using $\gamma = \gamma_r$ and $\gamma_r = \tilde{\gamma}_r$ (which we have shown in the previous subsection), yields $2\gamma_r = \mb{k}^\intercal\mb{d}^*$. Taking the complex conjugate of $2\gamma_r = \mb{k}^\intercal\mb{d}^*$ and combining it with $2\gamma_r = \mb{d}^\dagger \mb{d}$, we find:

\begin{equation}\tag{S4}
(\mb{\tilde{k}}^\dagger - \mb{\tilde{d}}^\dagger)\mb{d} = 0.
\end{equation}
As long as $\mb{d}$ is a non-zero vector, this implies $\mb{\tilde{k}} = \mb{\tilde{d}}$. Starting from the time-reversed scenario and following the same analysis, we can also confirm that $\mb{\tilde{d}} = \mb{\tilde{k}}$.

\subsection{Eq.~4e: $\mathbf{C}^\intercal\mb{d}^* = -\mb{k}$}\noindent
In the derivation for Eq.~4d we showed that $\mb{\tilde{C}}\mb{d}^*=-\mb{\tilde{d}}$. Combining this result with Eq. 4a, $\mathbf{\tilde{C}} = \mathbf{C}^\intercal$, and Eq. 4b,  $\mb{\tilde{d}} = \mb{k}$, we immediately find Eq.~4e. 

\subsection{Eq.~4f: $\mb{d}^\dagger \mb{d} = \mb{k}^\dagger \mb{k}$}\noindent
Finally, by using the fact that $\mb{C}$ is unitary, we obtain Eq.~4f from Eq.~4e: $\mb{d}^\dagger \mb{d} = \mb{k}^\dagger \mb{k}$. It is interesting to point out that there are various ways to derive this fluctuation-dissipation relation: one may also prove it using balance of power, or more rigorously, using stochastic methods \cite{Kampen1992}. Furthermore, it is important to point out that in the case of internal absorption, $\gamma_i>0$, the full fluctuation-dissipation relation needs to be amended to include absorptive dissipation as well.

In this context, we should stress that while these proofs rely on the unitarity of $\mb{C}$, the relations in Eq.~4 can also be more generally applied to lossy systems by considering loss as (an) additional port(s). This is specifically demonstrated by a heuristic derivation of Eq.~4e for the system with a dissipative wedge mode in Section 5 of the Supplementary Information. 

\section{Details of Cavity Geometries and Materials}\noindent
The unidirectional waveguides we use to study nonreciprocal cavities are based on the work by Shen et al. \cite{shen2015a,shen2015b}. The materials used in all of our simulations are the same: silicon (Si) and indium-antimonide (InSb). We use a constant permittivity for Si: $\varepsilon_{Si}=11.68 \varepsilon_0$, and for InSb we use the transversely ($\hat{z}$) magnetized permittivity tensor \cite{shen2015a}:

\begin{equation}\label{epsinsb1}\tag{S5}
\rttensor{\varepsilon}_{InSb} = \varepsilon_0 \varepsilon_\infty 
\begin{pmatrix}
 \varepsilon_1 & i \varepsilon_2 & 0 \\
  -i \varepsilon_2  & \varepsilon_1  & 0 \\
  0 & 0 & \varepsilon_3
\end{pmatrix}
\end{equation}
where $\varepsilon_\infty = 15.6 \varepsilon_0$ and 

\begin{subequations}\label{epsinsb2}
\begin{equation}\tag{S6a}
\varepsilon_1 = 1-\frac{(\omega - i \nu) \omega_p^2}{\omega ((\omega - i \nu)^2 - \omega_c^2)}
\end{equation}
\begin{equation}\tag{S6b}
\varepsilon_2 = \frac{\omega_c \omega_p^2}{\omega ((\omega - i \nu)^2 - \omega_c^2)}
\end{equation}
\begin{equation}\tag{S6c}
\varepsilon_3 = 1 - \frac{\omega_p^2}{\omega ((\omega - i \nu)^2 - \omega_c^2)}.
\end{equation}
\end{subequations}
Here $\omega_p = 4\pi \times 10^{12}$ rad/s is the plasma frequency, $\nu = 5 \times 10^{-3} \omega_p$ rad/s is the collision frequency, and $\omega_c = eB/m = -0.25 \omega_p$ is the cyclotron frequency (corresponding to a static magnetic field bias of 0.25 T in the $-\hat{z}$ direction).

In all simulations the waveguide has the same dimensions: the total height of the waveguide is 30 $\upmu$m, filled with 18 $\upmu$m of InSb at the bottom and 12 $\upmu$m of Si at the top. The bottom and top walls of the waveguide are perfect electric conductors (see the next section). For Fig.~1 in the main text we place a cavity behind the termination of the waveguide, which is 20 $\times$ 30 $\upmu$m and resonant at 1.52 THz. The cavity is connected to the waveguide through a small opening with height of 0.5 $\upmu$m and a width of 0.1 $\upmu$m. See Fig.~S1a for a schematic drawing of this geometry. We operate the waveguide in the unidirectional regime, with a pulse centered at 1.5 THz and a bandwidth of 0.16 THz (see next section).

\begin{figure*}[t]
\begin{center}
\includegraphics{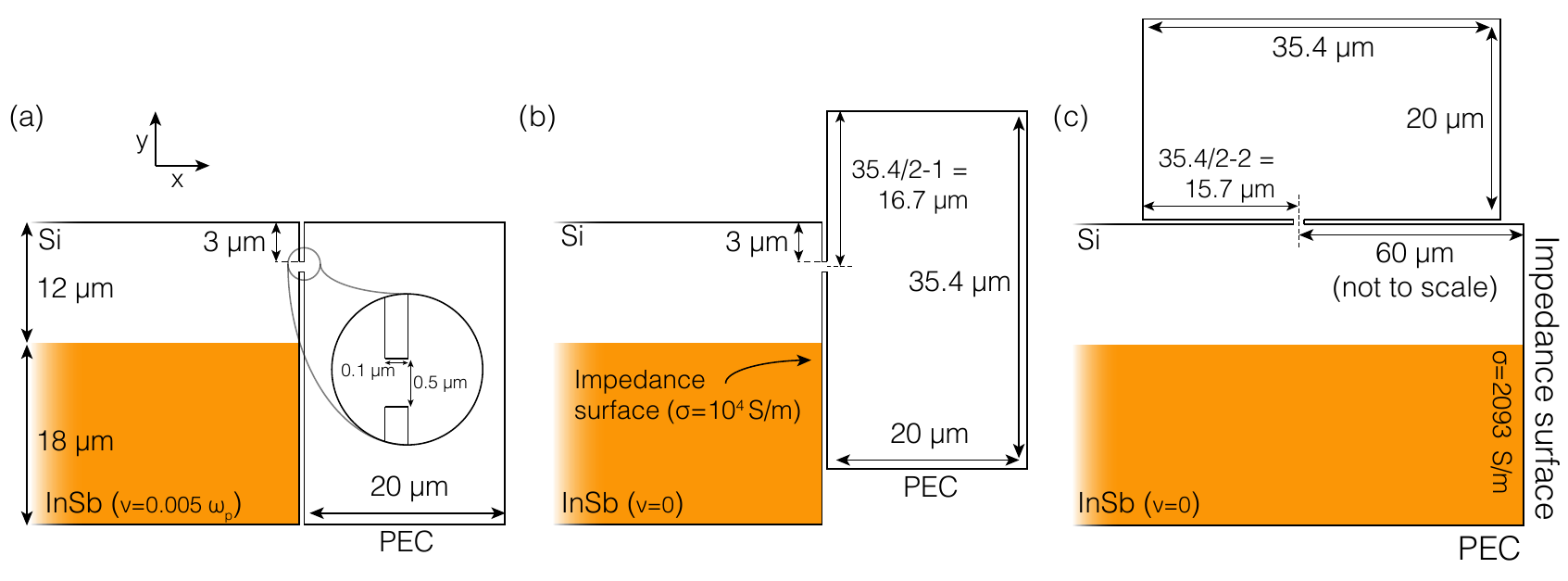}\renewcommand\thefigure{S1}
\caption{(a) Schematic for the first cavity discussed in the main text (and in Fig. 1), where the inset shows the geometry of the cavity opening. b) Schematic for the second system discussed in the main text (and in Fig. 2), where almost everything is identical as in the previous cavity except for a displacement of the cavity and a larger size. (c) The cavity is the same as in (b), except that now the cavity is on top and slightly displaced from the opening. In the waveguide, the opening is placed at a position where the forward and backward fields cancel out.} 
\label{fig1}
\end{center}
\end{figure*}

For Fig.~2 in the main text, we operate the waveguide in the bidirectional regime, below ~1.25 THz [3]. We increase the cavity size to 20 $\times$ 34.4 $\upmu$m so that it is resonant at 1.24 THz, and while we maintain the opening at the same position in the waveguide (3 $\upmu$m) from the top wall, we shift the cavity upwards so that the opening is closer to the middle of the cavity (which increases the cavity Q-factor). The displacement between the middle plane of the cavity and the middle of the opening is 2 $\upmu$m. For simplicity of analysis (see Section 4) we set the collision frequency to $\nu=0$ but have changed the PEC termination to a lossy impedance boundary condition with a conductivity of $\sigma = 10^4$ S/m and $\varepsilon = \varepsilon_0$. See Fig.~S1b for schematic details on this geometry.

The final geometry we have studied, discussed in Fig.~3 in the main text, has identical features to the previous cavity, except that the cavity is now positioned on top of the waveguide and that the conductivity of the termination has been reduced to 2093 S/m. The cavity opening is centered 60 $\upmu$m away from the termination, where there is a zero in the magnetic field (see Fig.~S1c).

\section{Anisotropic FDTD algorithm}\noindent
To perform full-wave simulations of the cavity in the time-domain we employ a home-built FDTD algorithm. For the geometries of interest, it is important for the algorithm to support i) Drude-model dispersion, and ii) anisotropic materials. To incorporate anisotropic Drude dispersion, we employ the auxiliary differential equation (ADE) method, which captures the dispersion in an additional equation for the current density \cite{Taflove2000} but implemented so that it supports anisotropic materials. Starting from Ampere's law in differential form:

\begin{equation}\tag{S7}
\nabla \times \mb{H} = \varepsilon_0 \frac{\partial \mb{E}}{\partial t} + \mb{J},
\end{equation}
where

\begin{equation}\tag{S8}
\mb{J} = i\omega \mb{P} = i\omega \varepsilon_0 \rttensor{\chi}_e \mb{E}.
\end{equation}
From the permittivity of InSb in Eqs.~\ref{epsinsb2}-\ref{epsinsb1}, we can write for the susceptibility

\begin{equation}\tag{S9}
\rttensor{\chi}_e = \frac{\varepsilon_\infty \omega_p^2}{\omega ((\omega-i\nu)^2-\omega_c^2)} 
\begin{pmatrix}
 -(\omega - i\nu) & i\omega_c \\
 -i\omega_c & -(\omega-i\nu)
 \end{pmatrix},
\end{equation}
so we find for the current density:

\begin{multline}\tag{S10}
\mb{J} = \frac{\varepsilon_0  \varepsilon_\infty \omega_p^2}{((\omega-i\nu)^2-\omega_c^2)} \\
\begin{pmatrix}
 -i(\omega-i\nu) & -\omega_c \\
 \omega_c & -i(\omega-i\nu)
 \end{pmatrix} \mb{E}.
\end{multline}
Inverting the matrix in this equation to bring it to the other side:

\begin{multline}\tag{S11}
\frac{-1}{((\omega-i\nu)^2-\omega_c^2)}
\begin{pmatrix}
 -i(\omega-i\nu) & \omega_c \\
 -\omega_c & -i(\omega-i\nu)
 \end{pmatrix}
 \mb{J} = \\
  \frac{\varepsilon_0  \varepsilon_\infty \omega_p^2}{((\omega-i\nu)^2-\omega_c^2)} \mb{E}
\end{multline}
Cancelling out the denominator, and using $i\omega\mb{J} \rightarrow \frac{\partial \mb{J}}{\partial t}$, we finally find:

\begin{equation}\tag{S12}
\frac{\partial \mb{J}}{\partial t} - \begin{pmatrix}
 \nu & \omega_c \\
 -\omega_c & \nu
 \end{pmatrix} \mb{J}
 = \frac{\partial \mb{J}}{\partial t} - \mb{\Omega}\mb{J} = \varepsilon_0  \varepsilon_\infty \omega_p^2 \mb{E}.
\end{equation}
This auxiliary differential equation incorporates the dispersion, and when converted into an update equation it can be added to a regular FDTD algorithm. We use the material parameters as discussed in the previous section, and a mesh of 10 by 10 nm.

\section{Heuristic Derivation of CMT for Cavity and Nonreciprocal Waveguide}\noindent
In the unidirectional waveguide discussed in Fig.~1 in the main text there is no reflected power over the whole frequency range of interest, and the direct reflection coefficient $C$ is therefore 0. We thus obtain an additional coefficient for the direct excitation of the wedge mode (which is responsible for absorbing the incident power), $\lvert C_w \rvert = 1 - \lvert C \rvert^2$, so that without the cavity the power absorbed by the wedge mode is $P_w = \lvert C_w s_+ \rvert^2$. While there is no backwards mode, even in the case of a lossless cavity as in Fig.~1 in the main text, the resonance can (and must be able to) dissipate power by exciting the wedge mode, which for the cavity results in an additional loss rate $\gamma_w$:

\begin{equation}\tag{S13}
\frac{da}{dt} - (i\omega_0 - \gamma_r - \gamma_i - \gamma_w)a +k_rs_+.
\end{equation}
It turns out that it is crucial to consider that both the incident wave and the resonance can excite the wedge mode, and that in due process they can thus interfere. The total power absorbed by the wedge mode in general is therefore given by

\begin{equation}\tag{S14}
P_w = \lvert C_w s_+ + d_w a\rvert^2
\end{equation}
where $d_w$ relates the excitation of the wedge mode to the cavity amplitude, with $\lvert d_w \rvert^2=2\gamma_w$ (which can be shown from balance of power in the case that $\gamma_r= \gamma_i=0$). If we then reconsider balance of power

\begin{equation}\tag{S15}
\lvert s_+ \rvert ^2 = \lvert s_ \rvert^2 + 2\gamma_i \lvert a \rvert^2 + P_w,
\end{equation}
expanding

\begin{equation}\tag{S16}
\lvert s_+ \rvert ^2 = \lvert C s_+ + d_r a\rvert^2 + 2\gamma_i \lvert a \rvert^2+ \lvert C_w s_+ + d_w a\rvert^2,
\end{equation}
which becomes 

\begin{multline}\tag{S17}
\lvert s_+ \rvert ^2 = (\lvert C \rvert^2 + \lvert C_w \rvert^2)\lvert s_+ \rvert ^2 +  2(\gamma_i + \gamma_r + \gamma_w) \lvert a \rvert^2 \\
+ 2\Re(C^*s_+^*d_ra) + 2\Re(C_w^* s_+^* d_w a).
\end{multline}
Inserting the equation for the cavity amplitude and setting $\omega = \omega_0$, we find:

\begin{multline}\tag{S18}
\frac{(\gamma_i + \gamma_r + \gamma_w) \lvert k_r s_+ \rvert^2}{(\gamma_i + \gamma_r + \gamma_w)^2} + \Re\biggl(\frac{C^*s_+^*dk_rs_+}{\gamma_i + \gamma_r + \gamma_w}\biggr) \\ 
+ \Re\biggl(\frac{C_w^*s_+^*d_wk_rs_+}{\gamma_i + \gamma_r + \gamma_w}\biggr),
\end{multline}
which becomes

\begin{equation}\tag{S19}
k_r^*k_r + \Re(C^*dk_r) + \Re(C_w^*d_wk_r)=0.
\end{equation}
Considering that $k_r^*k_r$ is real, we can write:

\begin{equation}\tag{S20}
-k_r^* = C_w^*d_w+C^*d.
\end{equation}
This equation is equivalent to Eq.~4d in the main text for this two-port system and it demonstrates that $k$ and $d$ can indeed be different, as also observed in the simulation, as long as there is an additional port. In the case that $d=0$ (when the waveguide is unidirectional), we again find that $\lvert k_r \rvert = \lvert d_w \rvert $, which means that power can enter the cavity at the same rate that the cavity can dissipate it via the wedge mode.

We stress again that it is crucial for the input and cavity to be able to interfere at the wedge. If this were not the case, and for example the wedge mode would dissipate incoherently as $P_w = \lvert C_w s_+ \rvert^2_+ \lvert d_w a \rvert^2 $, this would result again in the requirement $\lvert d_r \rvert= \lvert k_r \rvert$. In fact, in this scenario it is not necessary to consider the term $\gamma_w$ separately from $\gamma_i$, and a general description can be obtained without explicit consideration of $P_w$ and $\gamma_r$. This fact makes the wedge mode alike an additional channel, rather than simply an additional dissipative process.

\section{Using COMSOL for Nonreciprocal Waveguides}\noindent
For the results in Figs.~2-3 (as well as the intensity plots in Fig.~1) in the main text we use COMSOL rather than our FDTD algorithm, because we are interested in complex amplitudes of the ingoing and outgoing waves with respect to the cavity amplitude. In the following we describe how to use COMSOL for simulations with nonreciprocal media and the fitting procedure to obtain the coefficients discussed in the main text. For reciprocal media it would be most convenient to use COMSOL port boundary condition to launch and accept incoming and outgoing modes and to determine the reflection coefficient from the structure. However, COMSOL's numerical ports do not work with nonreciprocal waveguides, because the incoming and outgoing modes are different. We therefore applied the following procedure to obtain the complex mode and cavity amplitudes, assuming that we are modeling one of the structures in Fig.~S1:

\begin{figure*}[t]
\begin{center}
\includegraphics{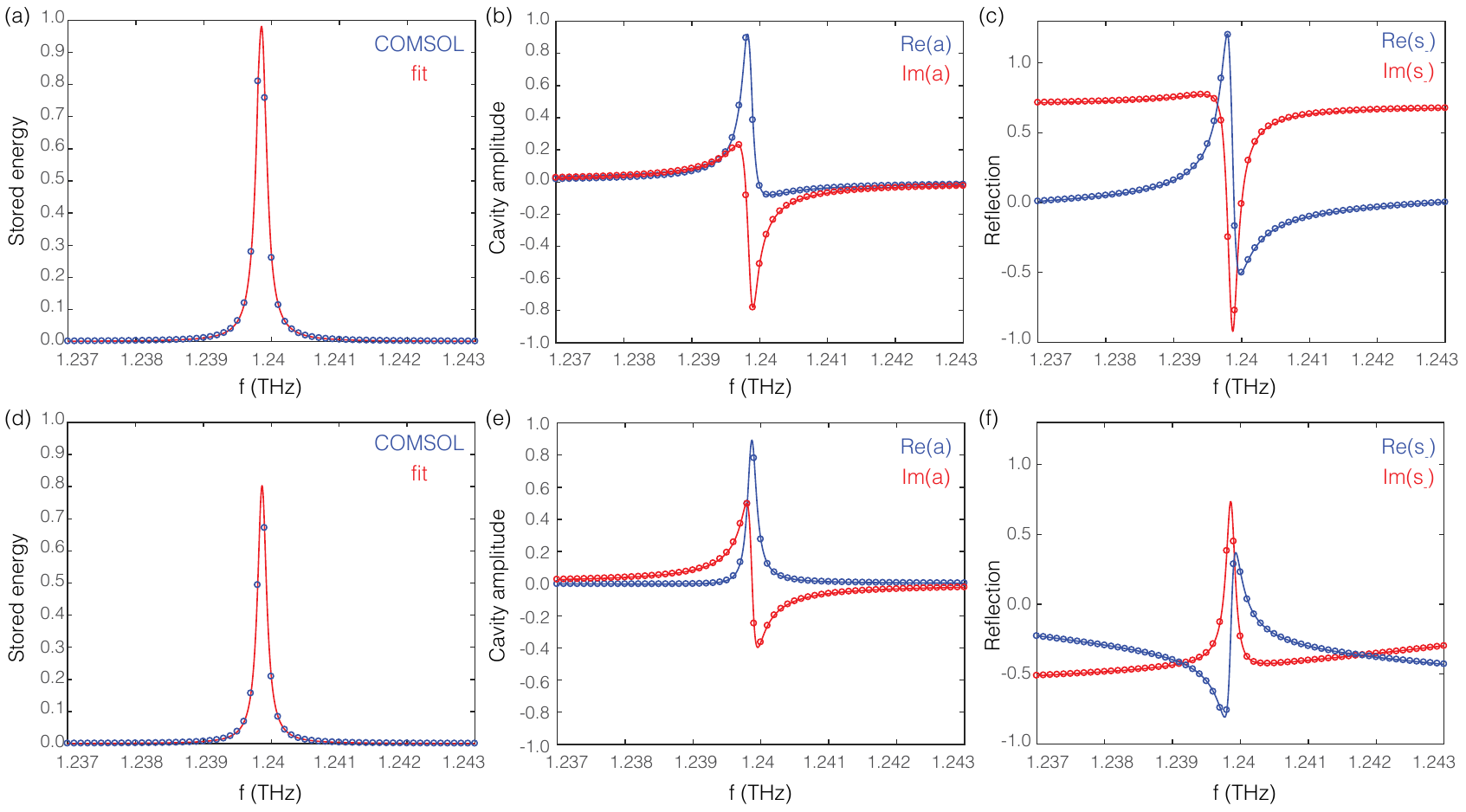}\renewcommand\thefigure{S2}
\caption{Fits of the stored energy (a), complex mode amplitude (b), and reflected amplitude (c) of a reciprocal cavity (without a magnetic bias). The bottom row of plots (d-e) shows the same, except for a nonreciprocal cavity (magnetic bias turned on). In all cases the fit is excellent, while fitting only one parameter (b,c,e,f). In (a,d) we fit three parameters: the loss rate, center frequency, and maximum stored energy (which we don't use in our analysis).} 
\label{fig2}
\end{center}
\end{figure*}

\begin{enumerate}
\item Create the geometry and add a port on the left end of the waveguide, with ``wave excitation'' set to ``on''.
\item To the left of where this port is, add a small second domain that is essentially an extension of the waveguide. Create a second \emph{Electromagnetic Waves, Frequency Domain} that applies only to this domain and make sure that in the main physics domain this waveguide extension is excluded.
\item In the second physics domain, set this waveguide extension up so that there is a port on the right boundary (overlapping with the port in the main physics domain). All other boundaries can be PEC.
\item Back in the first domain, add a second port on the same boundary, set it to a \emph{user defined} port and modify the expressions for the electric mode field and propagation constant so that they are obtained from the other physics domain: \emph{e.g.} ``$\texttt{emw2.tEmodex\_3}$'', \emph{etc}. Make sure wave excitation is set to ``off'' in this port. In this main physics domain there are now separate ports for ingoing and outgoing modes.
\item Although not strictly necessary (and slightly more complicated), it is possible to modify the weak expressions in the launching port so that it does not try to accept the returning mode. It requires adding an additional port for the electric field (following a similar procedure as before, but now in the same physics domain), and then add $0*$
 before all \texttt{if}'s in $\texttt{emw.PortConstrx}$, $\texttt{emw.PortConstry}$, $\texttt{emw.PortConstrz}$, $\texttt{emw.PortConstrx\_weak}$, $\texttt{emw.PortConstry\_weak}$, $\texttt{emw.PortConstrz\_weak}$, and the weak expression for the domain computation. 
\item Use an overlap integral on the input boundary to determine the input and output phase and amplitudes. Because these modes are nonreciprocal, the overlap integral is different \cite{mcisaac1991}:

\begin{equation}\tag{S21}
c_m = \frac{\iint (\mb{e}_{Tm} \times \mb{H}_T + \mb{E}_T \times \mb{h}_{Tm})\cdot \hat{z} dA}{2\iint \mb{e}_{Tm} \times \mb{h}_{Tm})\cdot \hat{z} dA}
\end{equation}
Here, $\mb{e}_{Tm}$ and $\mb{h}_{Tm}$ are the transverse mode profiles, while $\mb{E}_{T}$ and $\mb{H}_T$ are the transverse field profiles of the full solution.
\item Having obtained the input and output complex amplitudes, we can obtain the complex cavity amplitude from an integral over the stored electromagnetic energy in the cavity (due to the small opening and PEC walls virtually no energy is stored outside of the cavity) for the amplitude, and a field monitor in the center of the cavity for the phase.
\end{enumerate}

\section{Fitting Simulations with the Coupled-Mode Theory Model}

\noindent
We now describe the fitting procedure to obtain the complex coefficients reported in the main text. First, starting with the system shown in Fig.~\ref{fig1}b but without a magnetic field bias (so that it is reciprocal), we perform a frequency sweep for three different cavity sizes: one for the resonant cavity size ($20 \times 36.04$ $\upmu$m), and for cavity sizes one micron longer and shorter. We use these non-resonant cavities to determine the direct reflection path, by averaging $C=S_-/S_+$ for both simulations. Then, we obtain $\omega_0$ and $\gamma$ from a lorentzian fit of the stored energy in the cavity, which is shown in Fig.~\ref{fig2}a: $\omega_0=1.24$ THz and $\gamma$=96.7 MHz. To obtain $k_r$ we then proceed to the complex cavity amplitude, which we fit using 

\begin{equation}\tag{S22}
a=\frac{k_r s_+}{i(\omega-\omega_0)+\gamma}
\end{equation}
We use $\omega_0$ and $\gamma$ from the previous fit and $s_+$ is input from COMSOL, and hence \emph{only} $k_r$ is a fitting parameter. The fit of the complex cavity amplitude is shown in Fig.~\ref{fig2}b, for $k_r=(2.22-1.78) \times 10^4$ $\sqrt{\text{rad/s}}$. 

To obtain $d_r$ we fit the reflected amplitude:

\begin{equation}\tag{S23}
s_- = Cs_+ +d_r a
\end{equation}
Here, again, $d_r$ is the only fit parameter, and the rest we have obtained directly from COMSOL. Fig.~\ref{fig2}c shows the resulting fit, for $d_r=(222-1.79) \times 10^4$ $\sqrt{\text{rad/s}}$. As expected, $k_r=d_r$ but for a very small difference (the ratio is $d_r/k_r =1.0010-0.0007i$). The small differences most likely originates from estimating C (by changing the cavity size) and $a$ (by assuming that all of the stored energy is inside the cavity). The fact that we find $k_r=d_r$ in the reciprocal regime thus validates our method. If we now turn the magnetic field bias on again, we find the fits shown in Fig.~\ref{fig2}d-e, with $\omega_0=1.24$ THz and $\gamma=81.2$ MHz (indicating that the decay rate has changed, which is to be expected given the change in the waveguide mode field profiles). By fitting the complex cavity and reflection amplitudes we find $k_r=(2.65+0.308) \times 10^4$ $\sqrt{\text{rad/s}}$ and $d_r = (0.667+2.14) \times 10^4$ $\sqrt{\text{rad/s}}$. Now, clearly, the values are different: their ratio is $\frac{d_r}{k_r} = 0.34 +0.77 i$.

\bibliography{library.bib}